\documentclass[aps,floats]{revtex4}
\usepackage{amsmath,amssymb}
\usepackage{graphicx,epsfig}

\begin{document}
\bibliographystyle {plain}

\def\oppropto{\mathop{\propto}} 
\def\opsimeq{\mathop{\simeq}}
\def\opoverderline{\mathop{\overline}}
\def\operarrow{\mathop{\longrightarrow}}
\def\opsim{\mathop{\sim}} 
\def\opmin{\mathop{\min}} 
\def\opmax{\mathop{\max}} 
\def\oplim{\mathop{\lim}} 

\def\fig#1#2{\includegraphics[height=#1]{#2}}
\def\figx#1#2{\includegraphics[width=#1]{#2}}


\title{ Random cascade models of multifractality : \\  real-space renormalization and travelling-waves
} 


 \author{ C\'ecile Monthus and Thomas Garel }
  \affiliation{}
 \affiliation{ Institut de Physique Th\'{e}orique, CNRS and CEA Saclay,
 91191 Gif-sur-Yvette, France}

\begin{abstract}
Random multifractals occur in particular at critical points of disordered systems. For Anderson localization transitions, Mirlin and Evers [PRB 62,7920 (2000)] have proposed the following scenario (a) the Inverse Participation Ratios (I.P.R.) $Y_q(L)$ display the following fluctuations between the disordered samples of linear size $L$ : with respect to the typical value $Y^{typ}_q(L) = e^{\overline{\ln Y_q(L)}} \sim L^{-  \tau_{typ}(q)} $ that involve the typical multifractal spectrum $\tau_{typ}(q)$, the rescaled variable $y=Y_q(L)/Y^{typ}_q(L) $ is distributed with a scale-invariant distribution presenting the power-law tail $1/y^{1+\beta_q}$, so that the disorder-averaged I.P.R. $\overline{Y_q(L)} \sim L^{- \tau_{av}(q)} $ have multifractal exponents $\tau_{av}(q) $ that differ from the typical ones $\tau_{typ}(q)$ whenever $\beta_q<1$; (b) the tail exponents $\beta_q$ and the multifractal exponents are related by the relation $\beta_q \tau_{typ}(q)=\tau_{av}(q \beta_q)$. Here we show that this scenario can be understood by considering the real-space renormalization equations satisfied by the I.P.R. For the simplest multifractals described by random cascades, these renormalization equations are formally similar to the recursion relations for disordered models defined on Cayley trees and they admit travelling-wave solutions for the variable $(\ln Y_q)$ in the effective time $t_{eff}=\ln L$ : the exponent $\tau_{typ}(q)$ represents the velocity, whereas the tail exponent $\beta_q$ represents the usual exponential decay of the travelling-wave tail. In addition, we obtain that the relation (b) above can be obtained as a self-consistency condition from the self-similarity of the multifractal spectrum at all scales. Our conclusion is thus that the Mirlin-Evers scenario should apply to other types of random critical points, and even to random multifractals occurring in other fields.

\end{abstract}

\maketitle

\section{ Introduction }

To explain the motivations of the present work, we first need to recall
how the ideas of multifractality on one hand, and the idea 
of travelling waves on the other hand, 
have turned out to play a role in the field of disordered systems.

\subsection{ Multifractality and critical disordered systems in finite dimension }

\label{intro_multif}

Multifractality is a notion that has first appeared in fluid dynamics to
characterize the statistical properties of turbulence (see the book \cite{frisch} and references therein). Among the various areas 
where the multifractal formalism has then turned out to be relevant 
(see for instance \cite{halsey,Pal_Vul,Stan_Mea,Aha,Meakin,
harte,duplantier_houches} and references therein), 
the case of critical points in the presence of frozen
disorder is of particular interest.
The multifractal character of critical eigenfunctions at  
 quantum Anderson localization transitions has been the subject of 
very detailed studies (see the reviews \cite{janssenrevue,mirlinrevue}
and references therein).
More generally, multifractality of order parameters and correlation functions
 is expected to be a generic property of random critical points
whenever disorder is relevant :
it has been found in particular in disordered classical spin models like
random ferromagnets \cite{Ludwig,Jac_Car,Ols_You,Cha_Ber,Cha_Ber_Sh,PCBI,BCrevue}, spin-glasses or random field spin systems \cite{Sourlas,Thi_Hil,Par_Sou},
as well as in disordered polymer models like 
directed polymers in random media \cite{DPmultif}
or disordered wetting models \cite{wettingmultif}.
The only exceptions to these multifractal behaviors
seem to be  the ``multiscaling'' behaviors \cite{multiscaling},
which are even stronger than multifractality, that have been found
for some critical correlation functions
in disordered quantum spin-chains governed by 
 ``Infinite disorder fixed points'' \cite{revueigloi}.

\subsection{ Travelling waves and disordered systems }

Localized waves that propagate by keeping a fixed shape,
have been first discovered in fluid dynamics  in 1834
by J.S. Russel who wrote : `` that singular and beautiful phenomenon which I have called the Wave of Translation'' \cite{russel}.
The name `` Wave of Translation'' has not survived,
but the idea has flourished under other names.
 'Solitary waves' or 'solitons' have been found in many areas of physics
where non-linear equations of motion occur (see the book \cite{dauxois}
and references therein). 
 'Travelling waves' also appear in particular in the context of front
propagation into unstable states \cite{vanSaarloos} : 
in many cases called ``pulled fronts'', the velocity 
is actually determined by the exponentially small tail invading
 the unstable state, and can be thus determined by a linear analysis in the tail
region \cite{vanSaarloos,brunetreview}.
It turns out that this type of travelling waves also appear 
in the field of disordered systems, but with of course different meanings
for the space and time variables with respect to usual spatio-temporal waves.
It is useful to distinguish three cases :

(i) { \it In disordered models defined on Cayley trees} , it is the probability $P_L(A)$
of some observable $A$ that propagates without deformation 
in the effective 'time' corresponding to the length $L$ along the tree 
\begin{eqnarray}
t_{eff}=L
\label{teffL}
\end{eqnarray}
This property was discovered by Derrida and Spohn \cite{Der_Spohn}
on the specific example of the directed polymer in a random medium,
where the observable $A$ of interest is the free-energy,
and was then found in various other statistical physics models \cite{majumdar}.
This travelling-wave propagation of probability distributions
have also been found in quantum models defined on Cayley trees,
in particular in the Anderson localization problem
\cite{abou,zirnbauer,efetov,bell,us_cayley}
and in some superfluid-insulator transition \cite{ioffe_mez}.
The conclusion is thus that the recursion relations that can be written
for observables of disordered models defined on trees naturally lead to
the travelling wave propagation of the corresponding probability distributions. 
This property is not limited to the discrete Cayley trees, but actually still holds
for continuously branching trees \cite{Der_Spohn}.

(ii) { \it For some two-dimensional disordered models }, travelling waves
in the effective time given by the logarithm of the system size $L$
 \begin{eqnarray}
t_{eff}= \ln L
\label{tefflnL}
\end{eqnarray}
have been found, first for Dirac fermions in a random magnetic field 
by Chaman, Mudry and Wen \cite{mudry1,mudry2}, and then 
 for disordered XY models by Carpentier and Le Doussal \cite{carpentier_XY}, 
who  have derived non-linear
renormalization equations that admit travelling wave solutions. 
This approach has been then used to study other
 two-dimensional related models \cite{carpentier_log},
as well as the problem of a particle
in a logarithmically-correlated random Gaussian potential in finite dimension.
For this type of random energy models with logarithmically correlated potentials, many recent developments can be found in \cite{fyodorov_jpb,fyodorov,fyodorov_pld_ar}.

(iii) { \it In finite-dimensional Anderson localization models exactly at criticality} , 
where the eigenfunctions become multifractal (see the reviews \cite{janssenrevue,mirlinrevue} or section \ref{multif_loc} below), Evers and Mirlin 
\cite{Mirlin_Evers} have proposed
that the probability distribution $P_L( \ln Y_q)$ of the logarithm of the Inverse Participation Ratios $Y_q$ 
(which are the order parameters of the Anderson transition) propagate as travelling waves in the effective time $t_{eff}= \ln L$ given again 
by the logarithm of the system size $L$. 
This property has been checked numerically
in various Anderson localization models in various dimensions
\cite{Mirlin_Evers,Mirlin2002},
and has been recently obtained by a functional renormalization method in
dimension $d=2+\epsilon$ by Foster, Ryu and Ludwig \cite{foster_ludwig}.

In summary, travelling-wave propagation of probability distributions
have been found  
(i) in most disordered models defined on Cayley trees , 
(ii) in some specific two-dimensional disordered models or for a particle
in a logarithmically-correlated random potential, and 
(iii) at Anderson localization transitions in finite dimension.

\subsection{ Multifractality and travelling waves }

We believe that the case (iii) described above,
concerning Anderson localization transitions,
should actually apply to all random critical points in any finite dimension
that are characterized by multifractal properties.
For the directed polymer in a random medium of dimension $1+3$,
we have indeed found numerically the presence of travelling
waves at criticality \cite{DPmultif}. 
More generally, besides random critical points, we propose 
that random multifractal measures 
are generically related to the travelling wave propagation,
in the effective time $t=\ln L$, of the probability
distributions of the I.P.R. associated to the multifractal measure. 
Since the case of arbitrary multifractals is clearly beyond 
the scope of this paper, we will restrict our analysis
here to the simplest multifractal measures, namely the random cascade models
that have been much studied in the context of turbulence \cite{frisch}. 
These models satisfy simple real-space renormalization equations
that are formally similar to the recursion relations for disordered models
defined on Cayley trees (see case (i) described above). 
This interpretation thus allows
to understand the presence of travelling-wave at criticality in finite 
dimension, as a consequence of a hierarchical real space 
renormalization procedure on a appropriate
tree structure in the renormalization scale $\ln L$.

The paper is organized as follows.
In section \ref{travel_anderson}, we recall the multifractal notations and
 the Evers-Mirlin scenario concerning
the travelling waves that occur at Anderson localization transitions.
In section \ref{sec_rg}, we write real-space renormalization equations
to describe how the I.P.R. evolve upon coarse-graining.
In section \ref{sec_travelcascade}, we show that for random cascade models, 
these renormalization equations for the probability distributions
of the I.P.R. admit travelling-wave solutions. In section 
\ref{sec_relationtypavtail}, we obtain that 
the self-similarity of the multifractal spectrum at all scales
imposes the Mirlin-Evers relation $\beta_q \tau_{typ}(q)=\tau_{av}(q \beta_q)$
that relates the tail exponents $\beta_q$ of the travelling wave, to the typical and disorder-averaged multifractal spectra $(\tau_{typ}(q),\tau_{av}(q))$.
Our conclusions are summarized in section \ref{sec_conclusion}.

\section{ Reminder on the travelling waves at Anderson localization transitions }

\label{travel_anderson}

\subsection{ Reminder on typical and averaged multifractal spectra  }

\label{multif_loc}

At Anderson localization transitions, it is convenient to associate
to a critical eigenstate $\psi_L(\vec r)$ defined on a volume $L^d$
the measure
\begin{eqnarray}
\mu_L (\vec r) = \vert \psi_L(\vec r) \vert^2
\label{defmu}
\end{eqnarray}
normalized to 
\begin{eqnarray}
\int_{L^d} d^d { \vec r} \mu_L (\vec r) = 1
\label{norma}
\end{eqnarray}
The Inverse Participation Ratios (I.P.R.) of arbitrary order $q$ are defined by
\begin{equation}
Y_q(L)   \equiv \int_{L^d} d^d { \vec r}   \mu_L^q(\vec r)
\label{ipr}
\end{equation}
As a consequence of the normalization of Eq. \ref{norma},
one has the identity $Y_{q=1}(L)=1$.
The localization/delocalization transition can 
be characterized by the asymptotic behavior 
in the limit $L \to \infty$ of the $Y_q(L)$.
In the localized phase,
these moments $Y_q(L)$ converge to finite values
\begin{equation}
Y_q^{loc}(L=\infty) >0 
\label{loc}
\end{equation}
In the delocalized phase,  the decay
of the moments follows the scaling
\begin{equation}
Y_q^{deloc}(L) \oppropto_{L \to \infty} L^{ -(q-1) d } 
\label{deloc}
\end{equation}
Exactly at criticality, the typical decay of the $Y_q(L)$
defines a series of generalized exponents $\tau_{typ}(q)=(q-1) D_{typ}(q) $
\begin{equation}
Y_q^{typ}(L) \equiv e^{ \overline{ \ln Y_q(L)} }  
\oppropto_{L \to \infty} 
L^{ - \tau_{typ}(q) } = L^{ -(q-1) D_{typ} (q) } 
\label{tctyp}
\end{equation}
where the notation $\overline{A}$ denotes the average of the observable $A$
over the disordered samples.
The notion of multifractality corresponds to the case where $D_{typ}(q)$
depends on $q$, whereas monofractality corresponds to $D_{typ}(q)=cst$
as in Eq. (\ref{deloc}).
The exponents $D_{typ}(q)$ represent generalized dimensions \cite{halsey} :
$D_{typ}(0)$ represent the dimension of the support of the measure,
here it is simply given by the space dimension $D_{typ}(0)=d$;
 $D_{typ}(1)$ is usually called the information dimension \cite{halsey} ,
because it describes the behavior of 
the 'information' entropy
\begin{equation}
s_L \equiv - \sum_{ \vec r } \mu_L(\vec r) \ln  \mu_L(\vec r)
= - \partial_q Y_q(L) \vert_{q=1} \oppropto_{L \to \infty}  D_{typ}(1) \ln L
\label{entropy}
\end{equation}
Finally $D_{typ}(2)$ is called the correlation dimension \cite{halsey}
and describes the decay of
\begin{equation}
Y_2^{typ}(L) \equiv e^{ \overline{ \ln Y_2(L)} }   \oppropto_{L \to \infty} 
 L^{ - D_{typ} (2) } 
\label{y2d2}
\end{equation}

In the multifractal formalism, one also introduces
the singularity spectrum $f_{typ}(\alpha)$ defined as follows :
in a sample of size $L^d$, the number ${\cal N}_L(\alpha)$
of points $\vec r$ where the weight $\vert \psi_L(\vec r)\vert^2$
scales as $L^{-\alpha}$ behaves typically as 
\begin{eqnarray}
{\cal N}_L^{typ}(\alpha) \oppropto_{L \to \infty} L^{f_{typ}(\alpha)}
\label{nlalpha}
\end{eqnarray}
The saddle-point calculus in $\alpha$ of the I.P.R. 
\begin{equation}
Y_q^{typ}(L) \simeq \int d\alpha \ L^{f_{typ}(\alpha)} \ L^{- q \alpha} 
\label{saddle}
\end{equation}
yields the usual Legendre
transform formula 
\begin{eqnarray}
 -\tau_{typ}(q)  =  \opmax_{\alpha} \left[ f_{typ}(\alpha) - q \alpha  \right]
\label{legendre}
\end{eqnarray}

Following \cite{halsey}, many authors consider that
the singularity spectrum has a meaning only for $f_{typ}(\alpha) \geq 0$
\cite{Schreiber,Terao,Jan,Huck}. However, when
multifractality arises in random systems, disorder-averaged values
may involve other generalized exponents \cite{mandelbrot,Chh_neg,Jen_neg,has_dup}
than the typical values (see Eq. \ref{tctyp}), and it is thus useful
to introduce another series of generalized exponents   $\tau_{av}(q)=
(q-1) D_{av}(q) $  \cite{Mirlin_Evers}
\begin{equation}
\overline{ Y_q(L)}   \oppropto_{L \to \infty} 
 L^{ -\tau_{av}(q) } = L^{ -(q-1) D_{av} (q) } 
\label{tcav}
\end{equation}
For these disorder averaged values, the corresponding singularity
spectrum
$f_{av}(\alpha)$ may become negative $f_{av}(\alpha)<0$ 
\cite{mandelbrot,Chh_neg,Jen_neg,has_dup,Mirlin_Evers,vasquez}
to describe rare events (cf Eq. \ref{nlalpha}).
The difference between the two generalized exponents sets $D_{typ}(q)$ and
$D_{av}(q)$ associated to typical and averaged values 
has for origin the broad
distribution of the I.P.R. over the samples \cite{Mirlin_Evers,Mirlin2002}
as we now describe.

\subsection{ Probability distributions of the I.P.R. $ Y_q(L)$ over the samples }

\label{distriIPR}

The scenario proposed by Evers and Mirlin \cite{Mirlin_Evers,Mirlin2002} in the context of
quantum localization models 
is as follows : the probability distribution of the logarithm
of the Inverse Participation Ratios of Eq. \ref{ipr} becomes scale invariant
around its typical value
\cite{Mirlin_Evers,Mirlin2002}, i.e.
\begin{equation}
\ln Y_q(L) = \overline{\ln Y_q(L) } + u_q = \ln Y_q^{typ}(L) + u_q
\label{lnpq}
\end{equation}
where $u_q$ remains a random variable of order $O(1)$ 
in the limit $L \to \infty$. According to \cite{Mirlin_Evers}
the probability distribution
$G_L(u_q)$ generically develops an exponential tail
\begin{equation}
G_{\infty}(u_q) \oppropto_{u_q \to \infty} e^{ - \beta_q u_q}
\label{utail}
\end{equation}
As a consequence, the ratio $y_q =  Y_q(L)/Y_q^{typ}(L)=e^{u_q}$
with respect to the typical value $Y_q^{typ}(L)=e^{\overline{\ln Y_q(L)}}$
presents the power-law decay
\begin{equation}
\Pi \left( y_q \equiv \frac{Y_q(L)}{Y_q^{typ}(L)} \right) 
\oppropto_{ y_q \to \infty } \frac{1}{y_q^{1+\beta_q}}
\label{tailp}
\end{equation}
The conclusions of \cite{Mirlin_Evers,Mirlin2002} are then as follows :

(i) The typical singularity spectrum has a meaning only for $f_{typ}(\alpha) \geq 0$,
i.e.  it usually exists only on a finite interval $[\alpha_+,\alpha_-]$,
where the termination points $\alpha_{\pm}$ satisfy $f_{typ}(\alpha_-)=0=f_{typ}(\alpha_+)$.
Denoting $q_{\pm}$ the corresponding values of $q$ by the Legendre transformation
of Eq. \ref{legendre}, one obtains the linear behaviors outside the interval $[q_-,q_+]$
\begin{eqnarray}
\tau_{typ}(q) && = q \alpha_- \ \ {\rm for } \ \ q<q_- \\
\tau_{typ}(q) && = q \alpha_+ \ \ {\rm for } \ \ q>q_+
\label{lineartau}
\end{eqnarray}

(ii)
The disorder-averaged singularity spectrum
$f_{av}(\alpha)$ has a meaning outside this interval where $f_{av}(\alpha)<0 $
and describes the probabilities of rare events. In the region where both exist,
they are expected to coincide
\begin{equation}
f_{av}(\alpha)=f_{typ}(\alpha)  \ \ {\rm if } \ \ \alpha_+ \leq \alpha \leq \alpha_-
\label{coincidftildef}
\end{equation}
Equivalently, the disorder-averaged
exponents $\tau_{av}(q)$ are expected to coincide with the typical exponents
$\tau_{typ}(q)$ on the interval $[q_-,q_+]$
\begin{equation}
\tau_{av}(q)=\tau_{typ}(q)  \ \ {\rm if } \ \ q_- \leq q \leq q_+
\label{coincidtautildetau}
\end{equation}
whereas $\tau_{av}(q)$ will be different from Eq. \ref{lineartau}
outside the interval $[q_-,q_+]$.

(iii) from the point of view of the tail exponents $\beta_q$ of Eq. \ref{tailp},
this means that 
\begin{eqnarray}
 \beta_q &&>1  \ \ {\rm if } \ \ q_- < q < q_+   \\
 \beta_q &&<1  \ \ {\rm if } \ \ q<q_- \ \ {\rm or} \ \ q>q_+ \\
 \beta_{q_{\pm}}&& =1 
\label{xqcompareto1}
\end{eqnarray}

(iv) finally, Mirlin and Evers \cite{Mirlin_Evers} have derived in special cases the
following relation
\begin{eqnarray}
 \beta_q \tau_{typ}(q)=\tau_{av}(q \beta_q)
\label{relationtail}
\end{eqnarray}
and they have conjectured its generic validity \cite{Mirlin_Evers,mirlinrevue}.

This scenario has been tested numerically for various Anderson transitions
(see the review \cite{mirlinrevue} and the more recent works \cite{vasquez}).
We have found previously that this scenario also applies to
 the transition of the directed polymer in dimension $1+3$ \cite{DPmultif}.
In the following, we justify this scenario via a real-space renormalization analysis.

\section{ Real-space renormalization analysis  }

\label{sec_rg}

In critical phenomena, it is well known that critical properties
are stable under coarse-graining. This explains their
universal character (independence with respect to microscopic details)
and why renormalization is an appropriate framework.
Similarly for random critical points, the multifractal spectrum 
is expected to be stable under coarse-graining \cite{janssenrevue}.
It is thus natural to consider how the Inverse Participation Ratios (I.P.R.) evolve
upon coarse-graining.

\subsection{ Evolution of the I.P.R. upon coarse-graining   }

To go from the microscopic
 scale $l=1$ to the macroscopic scale $l=L$
of the whole system, it is convenient to introduce intermediate scales $l_m$
regularly placed on a logarithmic scale as follows.
For definiteness, we consider a discrete system with $L=b^M$ 
and introduce the intermediate scales 
\begin{eqnarray}
l_m=b^m \ \ {\rm with } \ \ m=0,1,..,M
\label{deflm}
\end{eqnarray}
Then the whole volume $L^d=l_M^d$ can be decomposed into 
\begin{eqnarray}
K \equiv b^d
\label{defK}
\end{eqnarray}
subvolumes of sizes $l_{M-1}^d$, and the process can be recursively iterated :
each volume of size $l_{m+1}^d= K l_m^d$ is made of
 $K$ volumes of sizes $l_m^d$, denoted here by an index $i=1,2,..K$.
We consider a (non-normalized) positive field $\mu(\vec r)$,
and associate to each volume $(i)$ of size $l_m$
the integrals of $\mu^q(\vec r)$ over the corresponding volume
\begin{eqnarray}
Z_q^{(i)}(m) \equiv \int_{l_m^d} d^d r \mu^{q}(\vec r)
\label{defZqm}
\end{eqnarray}
The corresponding I.P.R. are then defined by the ratios
\begin{eqnarray}
Y_q^{(i)}(m) \equiv \frac{Z_q^{(i)}(m)}{ \left[ Z_1^{(i)}(m) \right]^q}
\label{defYqm}
\end{eqnarray}

Upon coarse-graining, the integrals $Z_q$ are simply additive for any $q$
\begin{eqnarray}
Z_q(m+1) = \sum_{i=1}^{K} Z_q^{(i)}(m)
\label{RGZqm}
\end{eqnarray}
whereas the I.P.R. satisfy the recursions 
\begin{eqnarray}
Y_q(m+1) =  \sum_{i=1}^{K} \left[ w_i(m) \right]^q \ Y_q^{(i)}(m)
\label{RGqm}
\end{eqnarray}
where the coefficients are given by the weights
\begin{eqnarray}
w_i(m) \equiv 
 \frac{  Z_1^{(i)}(m) }{ \displaystyle \sum_{j=1}^{K} Z_1^{(j)}(m) }
\label{wqi}
\end{eqnarray}
that represent the ratios of the normalisation in the volume (i) of size $l_m^d$
with respect to the normalisation  of the volume of size $l_{m+1}^d$.
By construction one has the following constraint
\begin{eqnarray}
\sum_{i=1}^{K} w_i(m) =1
\label{normawqi}
\end{eqnarray}

For instance for Anderson localization models where the additive positive field
$\mu(\vec r)$ is given by Eq. \ref{defmu}, one expects that at sufficiently large scale

(i) in the delocalized phase, the $K$ weights $w_i$ all converge towards the same value $1/K$, so that the system becomes asymptotically 
homogeneous at sufficiently large scales.

(ii) in the localized phase, one single weight converge to $1$, whereas all other 
$(K-1)$ weights converge to zero.

(iii) exactly at criticality, the $K$ weights remain finite in contrast to (ii),
but they remain distributed with a non trivial distribution, in contrast to (i).

\subsection{ Notion of random cascade models }

Since the general case where the weights are correlated among generations
is more difficult to analyse, we will consider from now on
the much simpler case where the weights of different generations are {\it uncorrelated},
and where the non-negative weights $(w_1,...,w_{K})$
 corresponding to an elementary
coarse-graining step in Eq. \ref{RGqm} are drawn with some fixed
probability distribution $Q^*_b(w_1,...,w_{K})$, 
independent of the generation $m$, 
symmetric in its $K$ arguments, and satisfying
only the normalization constraint of Eq. \ref{normawqi} :
\begin{eqnarray}
Q^*_b(w_1,...,w_{K}) = R^*_b(w_1,...,w_{K}) \delta \left( \sum_{i=1}^{K} w_i(m) - 1 \right)
\label{fixedQ}
\end{eqnarray}
This type of random cascade models has been much studied in the context of turbulence to describe the spatial distribution of energy dissipation
(see the book \cite{frisch} and references therein).
In this context, the weights $(w_i)$ are usually called
'cascade generators' or 'multipliers' or 'breakdown coefficients'.
Various forms have been proposed over the years, in particular 
log-normal \cite{kolmogorov}, bimodal \cite{meneveau}, log-stable \cite{kida,schmitt}, log-Poisson \cite{she_lev,dubrulle,she}, log-infinitely-divisible \cite{novikov} : see \cite{cleve} for a comparative test of these various cascade generators.

However besides the specific form of the distribution of these weights,
the important hypothesis is of course the independence
 of the weights of different generations.
For turbulence, the validity of this hypothesis is
 discussed in \cite{sreen_sto,jouault}.
For random critical points, this hypothesis is not expected to be valid,
but one expects instead some Markovian structure \cite{multifsym}.
Nevertheless, since random cascade models are
 clearly the simplest multifractal models,
it is important to understand in detail their properties.
In the following, we show that travelling-waves appear 
very naturally in random cascade models, and that their properties are agree with
the Mirlin-Evers scenario concerning Anderson transitions.

\section{ Travelling-wave analysis for random cascade models }

\label{sec_travelcascade}

In the present section, we analyse the real-space renormalization
 Eq. \ref{RGqm}
for the case of random cascade models characterized by some fixed distribution
$Q^*_b(w_1,...,w_{K})$ (see Eq. \ref{fixedQ}). 
The renormalization Eq. \ref{RGqm} is then analogous to 
recursion equations for disordered models defined on a Cayley tree
of branching number $K$, and it is thus natural to obtain travelling-wave
propagation of probability distribution as for the directed polymer model
 \cite{Der_Spohn} mentioned in the introduction.
The fact that renormalization of multifractals in finite dimension
 involves a hierarchical structure
analogous to the Cayley tree has been already stressed in 
various contexts
\cite{mudry1,mudry2,carpentier_XY,carpentier_log,muzy}.

\subsection{ Travelling wave Ansatz 
for the Inverse Participation Ratios (I.P.R.)  }

We look for solutions of the real-space renormalization
 Eq. \ref{RGqm} with the following Ansatz
\begin{eqnarray}
Y_q(m) = e^{-v_q m} y_q
\label{travel}
\end{eqnarray}
where $v_q$ is a constant depending on $q$, and where $y_q$ is a random variable 
whose distribution does not depend upon the generation $m$.
In logarithmic scale, this corresponds to the travelling wave form
\begin{eqnarray}
\ln Y_q(m) = -v_q m+\ln  y_q
\label{logtravel}
\end{eqnarray}
where $v_q$ represents the velocity with respect to the variable
 $m= (\ln l_m)/(\ln b)$. The velocity $v_q$ is thus directly related
to the typical exponent $\tau_{typ}(q)$ of Eq. \ref{tctyp}
\begin{eqnarray}
v_q = (\ln b) \tau_{typ}(q) 
\label{logtravelconsistance}
\end{eqnarray}

With the Ansatz of Eq. \ref{travel}, the renormalization 
Eq. \ref{RGqm} becomes
\begin{eqnarray}
y_q = e^{v_q} \sum_{i=1}^{K} \left[ w_i \right]^q \ y_q^{(i)}
\label{yqrecurrence}
\end{eqnarray}
for the reduced random variables $y_q$.
The probability distribution $P^*_q(y_q)$ should be stable by this iteration
\begin{eqnarray}
P^*_q(y_q) = \int dw_1...dw_{K} Q^*_b(w_1,...,w_{K})
\int dy_q^{(1)} P^*_q(y_q^{(1)}) ... \int dy_q^{(K)} P^*_q(y_q^{(K)})
\delta \left( y_q - e^{v_q} \sum_{i=1}^{K} \left[ w_i \right]^q \ y_q^{(i)} \right)
\label{stability}
\end{eqnarray}

\subsection{ Tail analysis  }

A generic property of multiplicative stochastic processes
is to lead to probability distribution presenting power-law tails
 \cite{Kesten,Der_Pom,Bou,Der_Hil,Cal,us_quadratique}.
Here for the specific case of random multiplicative cascade models,
one expects also that the stable distribution $P^*_q(y_q)$
solution of Eq. \ref{stability} will present a power-law tail in $y_q$ 
\begin{eqnarray}
P^*_q(y_q) \opsimeq_{y_q \to \infty} \frac{ A}{y_q^{1+\beta_q} }
\label{tail}
\end{eqnarray}
where the exponent $\beta_q$ is not fixed for the moment,
and satisfies only the condition
$\beta_q>0$ to have a normalizable probability distribution.
In the travelling wave language of Eq. \ref{logtravel}, 
this power-law tail is equivalent to the exponential tail $e^{-\beta_q u_q}$
for the variable $u_q=\ln Y_q(m) +  m v_q=\ln y_q$.
So this corresponds exactly to the standard exponential tail 
analysis of travelling fronts \cite{vanSaarloos,brunetreview}.

The stability of the power-law tail of Eq \ref{tail}
in the region $y_q \to +\infty$ means that at leading order,
only one of the $K$ variable $y_q^{(i)}$ in Eq. \ref{stability}
becomes large with a probability also given by the tail of Eq. \ref{tail} :
after the introduction of a factor $K$ to choose one of the $K$ variables $y_q^{(i)}$, we may assume
the choice $i=1$ leading to 
\begin{eqnarray}
\frac{ A}{y_q^{1+\beta_q} }  && \simeq K
 \int dw_1...dw_{K} Q^*_b(w_1,...,w_{K})
\int dy_q^{(2)} P^*_q(y_q^{(2)}) ... \int dy_q^{(K)} P^*_q(y_q^{(K)})
\int dy_q^{(1)}  \frac{ A}{ (y_q^{(1)})^{1+\beta_q} }
\delta \left( y_q - e^{v_q}  \left[ w_1 \right]^q \ y_q^{(1)} \right)
\nonumber  \\
&& \simeq \frac{ A}{y_q^{1+\beta_q} } \  K 
 \int dw_1...dw_{K} Q^*_b(w_1,...,w_{K}) w_1^{q \beta_q} e^{ v_q \beta_q}
\label{tailstability}
\end{eqnarray}
One thus obtain the following compatibility equation
\begin{eqnarray}
1= K e^{v_q \beta_q } \overline{ w_i^{q \beta_q} }
\label{modebeta}
\end{eqnarray}
in terms of the partial moment of the joint probability distribution 
$Q^*_b(w_1,...,w_{K})$ of Eq. \ref{fixedQ}
\begin{eqnarray}
 \overline{ w_i^{p} } \equiv \int dw_1 dw_2 .. dw_K Q^*_b(w_1,...,w_{K})  w_i^p
\label{momentwi}
\end{eqnarray}

Eq. \ref{modebeta} means that 
each mode $\beta$ is characterized by the 
 velocity $v_q(\beta)$ given by
\begin{eqnarray}
v_q (\beta) = - \frac{1}{\beta} 
\ln  \left( K \overline{ w_i^{q \beta} } \right)
\label{vbeta}
\end{eqnarray}

\subsection{  Selection of the tail exponent $\beta_q$
 and of the velocity $v_q$ of the travelling wave }

In the field of travelling waves, 
the selection of the tail exponent $\beta$ of Eq. \ref{tail} 
and of the corresponding velocity $v(\beta)$
of Eq \ref{logtravel} usually depend on the form of the initial condition
\cite{Der_Spohn,vanSaarloos,brunetreview}. 
In our present case, the initial condition
  is completely localized $Y_q(m=0)=1$ at the lowest scale $l_{m=0}=1$.
One then expects that the solution that will
be dynamically selected \cite{Der_Spohn,vanSaarloos,brunetreview}
corresponds to the tail exponent $\beta^{selec}_q$
and to the velocity $v^{selec}_q=v_q(\beta^{selec}_q)$
determined by the following extremization criterion
\begin{eqnarray}
0= \left[ \partial_{\beta} v_q(\beta) \right]_{\beta=\beta^{selec}_q}
= \left[ \frac{1}{\beta^2} \ln   \left( K \overline{ w_i^{q \beta} } \right) 
- \frac{1}{\beta}  \partial_{\beta} \ln   \left(  \overline{ w_i^{q \beta} } \right) 
\right]_{\beta=\beta^{selec}_q}
\label{selection}
\end{eqnarray}

In summary, for any given distribution $Q^*_b(w_1,...,w_{K})$
 that defines a random cascade
model (see Eq. \ref{fixedQ}), the properties of the travelling waves
can be obtained as follows : 
one has to compute the partial moment of Eq. 
\ref{momentwi}, and to solve Eq. \ref{selection} in order to determine the
tail exponent $\beta_q$ and the velocity $v_q$ that are dynamically selected.
The typical multifractal exponents $\tau_{typ}(q)$ are then obtained 
from the selected velocities by Eq. \ref{logtravelconsistance}
\begin{eqnarray}
 \tau_{typ}(q) = \frac{v_q^{selec}}{(\ln b)}
\label{tautypselec}
\end{eqnarray}

\section{ Relations between the tail exponents and the multifractal exponents  }

\label{sec_relationtypavtail}

As explained in the previous section,
 the selection criterion of Eq. \ref{selection} is usually the final outcome
 of a travelling wave analysis.
However here in our real-space renormalization framework,
we still have some freedom in the choice of the rescaling factor $b$
introduced in Eq. \ref{deflm}. Of course, the final multifractal exponents 
$\tau_{typ}(q),\tau_{av}(q)$ should not depend on the choice of the coarse-graining scale $b$.
In the present section, we use this freedom to consider the case
of large $b$, and we show that some self-consistency conditions arise.

\subsection{ Self-consistency conditions 
when the rescaling factor $b$ is large  }

When the rescaling factor $b$ introduced in Eq. \ref{deflm} is large,
the number $K=b^d$ of buildings blocks (see Eq. \ref{defK}) 
of a single renormalization step
also becomes large. Then from the self-similarity of the multifractal spectrum
at all scales \cite{janssenrevue}, one obtains that the
elementary weights $w_i$ should themselves follow
the multifractal statistics, i.e. their associated I.P.R. 
\begin{eqnarray}
{\cal Y}_p (b) \equiv
 \sum_{i=1}^{K=b^d} w_i^p 
\label{wiKlarge}
\end{eqnarray}
should 
have for typical scalings
\begin{eqnarray}
{\cal Y}_p^{typ} (b) \oppropto_{b \to +\infty}  b^{-\tau_{typ}(p)}
\label{typwiKlarge}
\end{eqnarray}
and for averaged scalings 
\begin{eqnarray}
\overline{ {\cal Y}_p } (b) =
K \overline{ w_i^p } \oppropto_{b \to +\infty}   b^{-\tau_{av}(p)}
\label{moywiKlarge}
\end{eqnarray}
in terms of the typical and averaged multifractals exponents $\tau_{typ}(p)$ and $\tau_{av}(p)$ introduced in Eqs \ref{tctyp} and \ref{tcav}.
The velocity of Eq. \ref{vbeta} then reads at leading order for large $b$
\begin{eqnarray}
v_q (\beta) = - \frac{1}{\beta} 
\ln  \left( b^{- \tau_{av}(p=q \beta )} \right)
= \frac{\tau_{av}(p=q \beta)}{ \beta } \ln b 
\label{vbetanew}
\end{eqnarray}
Taking into account Eq. \ref{tautypselec}, 
we thus obtain the following consistency relation
\begin{eqnarray}
 \tau_{typ}(q) = \frac{v_q^{selec}}{(\ln b)} = \frac{\tau_{av}(p=q \beta_q^{selec})}{ \beta_q^{selec} }
\label{relationtypavbeta}
\end{eqnarray}
that relates the typical exponent $\tau_{typ}(q)$
 to the selected tail exponent 
$\beta_q^{selec}$ and to the averaged multifractal exponent 
$\tau_{av}(p=q \beta_q^{selec})$.
As recalled around Eq. \ref{relationtail}, the relation
of Eq. \ref{relationtypavbeta} has already been derived in special cases
by Mirlin and Evers \cite{Mirlin_Evers} and has been conjectured to be general 
\cite{mirlinrevue}. Our present derivation from a self-consistency condition
of the travelling wave analysis is thus in favor of 
 the general validity of this formula.

The relations of Eq. \ref{relationtypavbeta} introduces non-trivial constraints
on the multifractal exponents, which are not always compatible with
the usual selection criterion of Eq. \ref{selection}
based on the extremization of the velocity.
Let us first describe an explicit case before returning to the general case.

\subsection{ Example of Gaussian multifractality }

The simplest multifractal spectrum corresponds to the following Gaussian
form for the disorder-averaged multifractal spectrum
\begin{eqnarray}
\tau_{av}^{Gauss}(q) = d (q-1) \left( 1- \frac{q}{q_c^2} \right)
\label{tauavGauu}
\end{eqnarray}
In particular, this Gaussian forms appears in various models
in the weak multifractality regime \cite{mirlinrevue}, 
in particular in perturbation theory in $d=2+\epsilon$ \cite{foster_ludwig}.

\subsubsection{ Consequences of the self-consistency Eq. \ref{relationtypavbeta} }

In the region where $\tau_{typ}^{Gauss}(q)=\tau_{av}^{Gauss}(q)$ 
corresponding to $\beta_q>1$,
Equation \ref{relationtypavbeta} yields \cite{mirlinrevue}
\begin{eqnarray}
\beta_q = \frac{q_c^2}{q^2}
\label{betaGaussregiontyp}
\end{eqnarray}
This solution is consistent for $\beta_q>1$, i.e. in the interval
\begin{eqnarray}
-q_c < q < +q_c 
\label{consistanceGaussregiontyp}
\end{eqnarray}

In the region where $\tau_{typ}^{Gauss}(q) \ne \tau_{av}^{Gauss}(q)$ 
corresponding to $\beta_q<1$,
the typical exponents vary linearly with $q$ \cite{mirlinrevue,foster_ludwig}
\begin{eqnarray}
\tau_{typ}^{Gauss}(q) = q d  \left( 1- \frac{  {\rm sgn}(q) }{q_c} \right)^2  {\rm for } \ \ \vert q \vert > q_c  
\label{tautypGaussregionlin}
\end{eqnarray}
Equation \ref{relationtypavbeta} then yields \cite{mirlinrevue,foster_ludwig}
\begin{eqnarray}
\beta_q = \frac{q_c}{\vert q \vert}
\label{betaGaussregionrare}
\end{eqnarray}

\subsubsection{ Analysis via the selection criterion of Eq. \ref{relationtypavbeta} }

From Eq. \ref{tauavGauu}, we obtain the velocity as a function of
 the tail exponent $\beta$ using Eq. \ref{vbetanew}
\begin{eqnarray}
v_q (\beta) 
= \frac{\tau_{av}(p=q \beta)}{ \beta } \ln b 
=  \frac{(\ln b) }{ \beta }   d (\beta q-1) \left( 1- \frac{\beta q}{q_c^2} \right)
\label{vbetagauss}
\end{eqnarray}
Its derivative
\begin{eqnarray}
\partial_{\beta} v_q (\beta) = d (\ln b) \left[ \frac{1}{\beta^2} - \frac{q^2}{q_c^2} \right]
\label{partialvbetagauss}
\end{eqnarray}
yields the following solution $\beta^{selec}_q>0$
for the selection criterion of Eq. \ref{selection} 
\begin{eqnarray}
\beta^{selec}_q = \frac{q_c}{\vert q \vert} 
\label{betaselecgauss}
\end{eqnarray}
and the corresponding selected velocity reads
\begin{eqnarray}
v_q^{selec} =v_q (\beta_q^{selec})  
= (\ln b) d  q  
 \left( 1- \frac{  {\rm sgn}(q) }{q_c} \right)^2
\label{vselecgauss}
\end{eqnarray}
The typical multifractal exponent reads  (Eq. \ref{tautypselec})
\begin{eqnarray}
 \tau_{typ}(q) = \frac{v_q^{selec}}{(\ln b)}
=  d  q  
 \left( 1- \frac{  {\rm sgn}(q) }{q_c} \right)^2
\label{tautypselecgauss}
\end{eqnarray}
i.e. one recovers the correct result
 {\it only in the linear regions of the typical
spectrum, where it is different from the disorder-averaged spectrum }.
This seems to indicate that in the region where $\tau_{typ}(q)=\tau_{av}(q)$,
this additional constraint $\tau_{typ}(q)=\tau_{av}(q)$
fixes completely the tail exponent and the velocity to values
that do not satisfy the usual
 selection criterion of Eq. \ref{relationtypavbeta}.
This phenomenon seems generic even beyond the Gaussian case as we now explain.

\subsection{ General case : competition with the usual velocity selection  }

For the general case, the selection condition of Eq. \ref{selection}
 for $\beta^{selec}_q$ becomes
\begin{eqnarray}
0 = \left[ q \beta \tau_{av}'(q \beta )
 - \tau_{av}(q \beta ) \right]_{\beta^{selec}_q}
\label{selectionnew}
\end{eqnarray}
In terms of the solution $p^*$ of the following equation
\begin{eqnarray}
0 = \left[ p \tau_{av}'( p )
 - \tau_{av}( p ) \right]_{p=p^*}
\label{selectionp}
\end{eqnarray}
the selected tail exponent reads
\begin{eqnarray}
\beta^{selec}_q =  \left\vert \frac{p^*}{ q  } \right\vert
\label{betaselecq}
\end{eqnarray}
and the corresponding velocity is given by
\begin{eqnarray}
v_q^{selec} = v_q (\beta^{selec}_q) 
= \frac{\tau_{av}(q \beta^{selec}_q)}{ \beta^{selec}_q } \ln b
\label{vselecnew}
\end{eqnarray}
Eq. \ref{logtravelconsistance} 
yields the following typical exponent 
\begin{eqnarray}
\tau_{typ}(q) = \frac{v_q^{selec}}{\ln b} = \
\frac{\tau_{av}(q \beta^{selec}_q)}{ \beta^{selec}_q }
= q \frac{\tau_{av}(\vert p^*\vert  {\rm sgn}(q) )}{ \vert p^* \vert {\rm sgn}(q) }
\label{logtravelconsistancebis}
\end{eqnarray}
which is linear in $q$.

In summary, as in the Gaussian case, we obtain that 

(i) the usual selection criterion
of Eq. \ref{selection} is compatible with the self-consistency 
Eq. \ref{relationtypavbeta}
only in the region where the typical and averaged spectra differ
 $\tau_{typ}(q) \ne \tau_{av}(q)$ where $\beta_q<1$. 

(ii) In the region where $\tau_{typ}(q) = \tau_{av}(q)$ and $\beta_q>1$,
the self-consistency Eq. \ref{relationtypavbeta} indicates that 
the selected tail exponent is the solution $\beta_q^{selec}>1$ of
\begin{eqnarray}
 \beta_q^{selec} \tau_{av}(q) = \tau_{av}(q \beta_q^{selec})
\label{relationavav}
\end{eqnarray}
(the other trivial solution of this equation being $\beta=1$),
and does not correspond to the usual selection criterion
of Eq. \ref{selection}.

\section{ Conclusions and Perspectives }

\label{sec_conclusion}

In this paper, we have proposed that the Mirlin-Evers scenario
concerning the probability distributions of Inverse Participation Ratios (I.P.R.)
at Anderson transitions can be better understood by
considering the real-space renormalization equations satisfied
by I.P.R. upon coarse-graining.
 For the simplest multifractal models, namely the random cascade models, we have shown that these renormalization equations are formally similar
to the recursions equations for disordered models defined on Cayley trees,
and that, as a consequence, they admit travelling wave solutions
of the type known as ``pulled fronts'', where the velocity 
is actually determined by a linear analysis in 
the exponentially-small tail region.
Finally, we have obtained that 
the self-similarity of the multifractal spectrum at all scales
imposes the Mirlin-Evers relation of Eq. \ref{relationtail}
that relates the tail exponents $\beta_q$ of the travelling waves, to the typical and disorder-averaged multifractal spectra $(\tau_{typ}(q),\tau_{av}(q))$.
This shows that random cascade models already capture many properties 
that have been previously found at Anderson transitions.
It would be thus interesting in the future 
to show that the travelling wave solutions
of the renormalization equations persist beyond random cascade models,
since at random critical points, one expects that the strict 
statistical independence
 of the weights of different generations is not valid,
but one expects instead some Markovian structure \cite{multifsym}.
Further work is needed to formulate correctly appropriate models of
Markovian cascades.

Nevertheless, since random cascade models are
 clearly the simplest multifractal models,
and since their properties are very similar to the properties found previously for the more complex case of Anderson transitions, we believe that
our present real-space renormalization analysis 
is in favor of a wide validity of the Mirlin-Evers scenario
beyond Anderson transitions :

(i) the first generalization concerns all random critical points in finite dimension, where the local order parameter and the correlation functions 
generically display multifractal statistics (see the discussion in section 
\ref{intro_multif} of the Introduction).
For instance, we have found numerically the presence of travelling
waves at criticality for the directed polymer in a random medium of dimension $1+3$,
 \cite{DPmultif}. For other many-body random phase transitions
like classical disordered spin models, these travelling wave properties
should also appear after an appropriate translation (see \cite{multifsym}
for such a translation between multifractality at Anderson transitions
and in classical disordered spin models). 
It seems that for two-dimensional disordered Potts models where multifractality
has been much studied \cite{Ludwig,Jac_Car,Ols_You,Cha_Ber,Cha_Ber_Sh,PCBI,BCrevue,multifsym}
only the disorder-averaged multifractal spectrum
 has been considered up to now.
It would thus be very interesting to study numerically the
probability distribution over the disordered samples of the appropriate observables
to test the travelling-wave scenario and to measure
the corresponding tail exponents $\beta_q$.
Some probability distributions over the disordered samples
of extensive observables like the susceptibility have been already measured
for some random critical points in \cite{domany95,domany,berche,chatelain},
but a quantitative analysis of the tails remains to be done.

(ii) besides phase transitions in disordered systems,
 a further generalization concerns other areas of physics
where multifractality occurs. We believe 
that the present real-space renormalization analysis on the multifractal measure keeps its validity,
provided the notion of fluctuations between different realizations of
the multifractal cascade has a physical meaning.
Indeed, in disordered systems, each disordered sample
is characterized by a given realization of the multifractal cascade,
so that the fluctuations between cascade realizations correspond to
the sample-to-sample fluctuations, which play in major role in the understanding
of disordered systems. In other fields where multifractality occurs, 
one should first clarify the physical meaning of a given realization of the cascade to see whether it is interesting to distinguish between typical exponents and disorder-averaged exponents, and to introduce probability distributions over 
cascade realizations.

\section*{Acknowledgements}

It is a pleasure to thank Olivier Giraud for a very useful discussion on 
multifractality and tail exponents.

\end{document}